\documentclass[twocolumn]{aastex631}
\usepackage{graphicx}
\usepackage{booktabs}
\usepackage{multirow}
\usepackage{gensymb}
\usepackage{float}
\usepackage{subfigure}

\newcommand\lsim{\mathrel{\rlap{\lower4pt\hbox{\hskip1pt$\sim$}}
\raise1pt\hbox{$<$}}}
\newcommand\gsim{\mathrel{\rlap{\lower4pt\hbox{\hskip1pt$\sim$}}
\raise1pt\hbox{$>$}}}
\begin{document}

\title{Upper Limits on Stellar Companions to the Kepler-34 and Kepler-35 Systems}

\correspondingauthor{Carlos Jurado}
\email{carx207@g.ucla.edu}
\author[0009-0009-7568-8851]{Carlos Jurado}
\affiliation{Department of Physics and Astronomy, University of California, Los Angeles, Los Angeles, CA 90095, USA}
\affiliation{Department of Physics and Astronomy, University of Notre Dame, Notre Dame, IN 46556, USA}

\author[0000-0002-3725-3058]{Lauren M. Weiss}
\affil{Department of Physics and Astronomy, University of Notre Dame, Notre Dame, IN 46556, USA}

\author{Laura Daclison}
\affiliation{Institute for Astronomy, University of Hawaii at Manoa, Honolulu, HI 96822, USA}

\author[0000-0003-2053-0749]{Benjamin M.\ Tofflemire}
\altaffiliation{51 Pegasi b Fellow}
\affiliation{Department of Astronomy, The University of Texas at Austin, Austin, TX 78712, USA}

\author[0000-0001-9647-2886]{Jerome A. Orosz}
\affiliation{Department of Astronomy, San Diego State University, San Diego, CA 92182, USA}

\author[0000-0003-2381-5301]{William F. Welsh}
\affiliation{Department of Astronomy, San Diego State University, San Diego, CA 92182, USA}

\begin{abstract}
We obtained new spectra of Kepler-34 and Kepler-35 with Keck-HIRES, nearly a decade after these systems were originally characterized with this spectrograph and other instruments, to search for RV trends from a potential third stellar-mass companion at long periods. For Kepler-34, we rule out coplanar stellar masses as low as $0.12 M_\odot$ at an orbital period of $\lesssim 52$ years. For Kepler-35, we rule out stellar masses of $0.13 M_\odot$ at orbital periods of $\lesssim 55$ years. Highly stable, extreme precision RV instruments, as well as improved methodologies in characterizing double-lined spectroscopic binaries that come with these new instruments, will provide an opportunity to push these mass limits lower in the future.

\end{abstract}

\section{Introduction}
\label{sec:intro}
Circumbinary planets orbit around both stars of a stellar system. The first confirmed detection of a transiting circumbinary planet, Kepler-16b \citep{Doyle2011}, was based on photometry from the NASA \textit{Kepler} mission. Since then, a total of 14 circumbinary planets have been detected from \textit{Kepler} photometry. Notably, the transiting circumbinary planets have provided the majority of circumbinary planet detections that form the basis of our understanding of their demographics \citep{Armstrong2014, Martin2014, Li2016}.



\begin{figure*}
    \centering
    \subfigure[2011-09-02 08:33:24 UT]{%
        \includegraphics[width=0.3\textwidth]{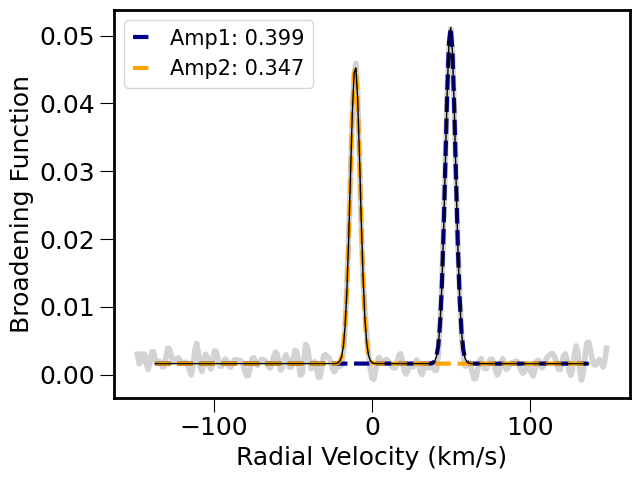}
        \label{fig:image1}
    }
    \hfill
    \subfigure[2011-09-05 11:58:33 UT]{%
        \includegraphics[width=0.3\textwidth]{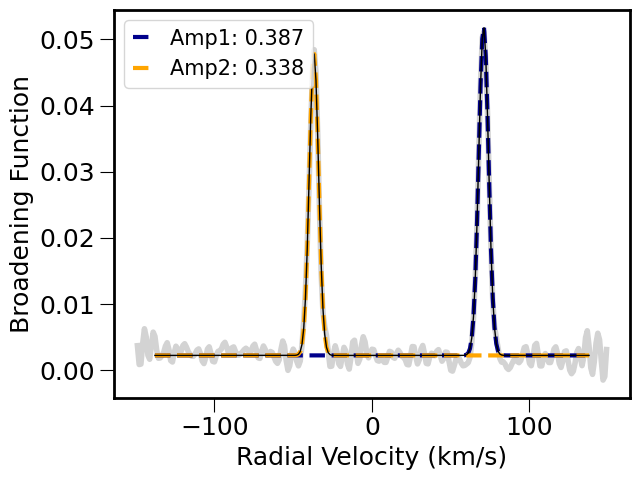}
        \label{fig:image2}
    }
    \hfill
    \subfigure[2011-09-06 11:47:10 UT]{%
        \includegraphics[width=0.3\textwidth]{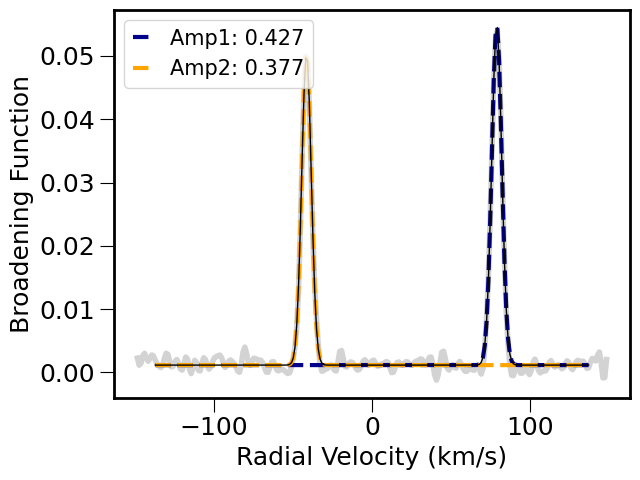}
        \label{fig:image3}
    }
    
    \vspace{0.5cm}
    
    \subfigure[2011-09-10 07:48:37 UT]{%
        \includegraphics[width=0.3\textwidth]{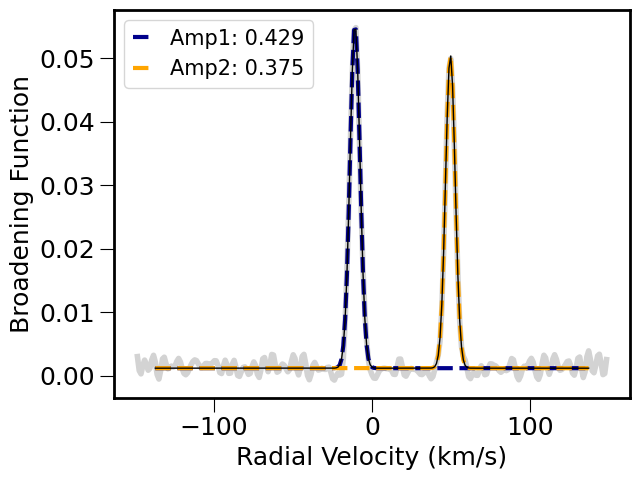}
        \label{fig:image4}
    }
    \hfill
    \subfigure[2019-12-15 05:16:40 UT]{%
        \includegraphics[width=0.3\textwidth]{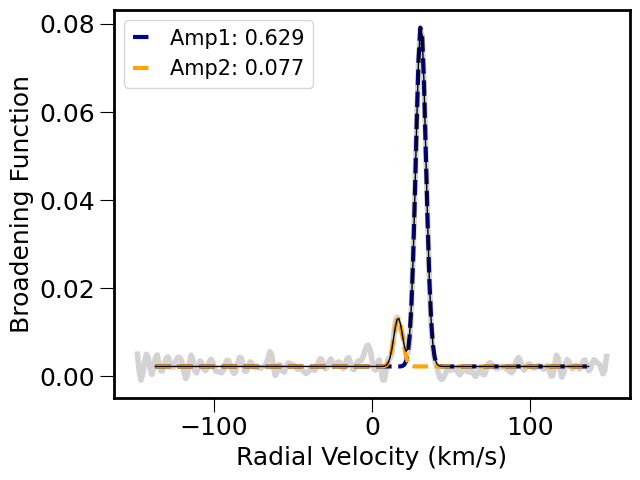}
        \label{fig:image5}
    }
    \hfill
    \subfigure[2021-08-30 10:13:38 UT]{%
        \includegraphics[width=0.3\textwidth]{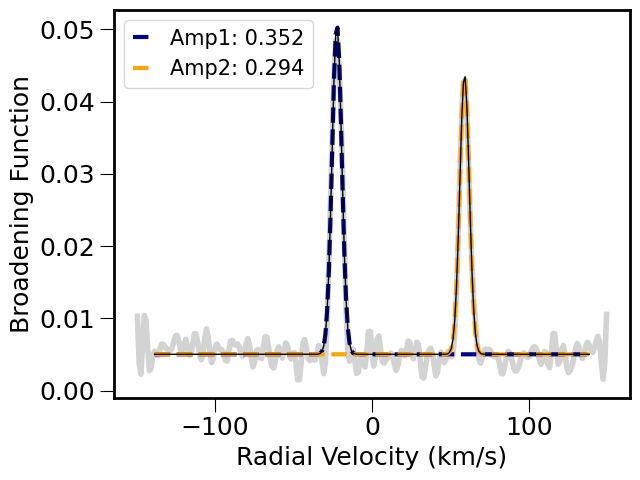}
        \label{fig:image6}
    }
    
    \caption{Broadening Function of HIRES spectra of Kepler-34 obtained in six unique observations. Gaussian profiles are fitted to the peaks in the Broadening Function at the radial velocities associated with the primary star (blue) and secondary star (orange). The area underneath the Gaussian profiles for each star are provided in the legend.}
    \label{fig:broadfunc_kep_34}
\end{figure*}


Radial velocities (RVs) have the potential to further our understanding of circumbinary planets. RV curves of double-lined spectra allow for a detailed characterization of the orbital properties and masses of both the primary and secondary star. In principle, with sufficient RV precision, the planetary signal in RV curves of circumbinary host stars directly probe the planet's mass. However, in practice, this type of measurement is exceptionally challenging, particularly for double-lined spectroscopic binaries.  Although the RV method has proven to be a highly effective way to detect planets around single stars, it has been challenging to obtain the radial velocity precision necessary to detect circumbinary planets. The spectra of double-lined binaries are both composite and time-variable and thus, traditional methods of radial velocity determination with an iodine absorption cell cannot be applied \citep{Marcy1992}. Previous radial velocity searches of double-lined spectroscopic binaries, \citep[TATOOINE;][]{Konacki2009, Konacki2010} were able to rule out regions of mass-period parameter space of circumbinary planets but did not yield any detections. Recently, spectroscopic studies targeting single-lined, low-mass eclipsing binaries, such as those conducted by BEBOP \citep{Martin2019+}, have led to the radial velocity detection of Kepler-16 b \citep{Triaud2022} and TOI-1338/BEBOP-1 c \citep{Standing2023}.


Nonetheless, long-term RV monitoring of double-lined spectroscopic binary planet hosts can provide unique constraints on the stellar and planetary system architectures. In this work, we revisit the Kepler-34 and Kepler-35 systems, both of which are double-lined spectroscopic binaries with a transiting planet \citet{Welsh+12}.  Our motivation to return to these systems is to establish a decade-long baseline with additional RV measurements from Keck-HIRES.  We re-reduce and analyze all existing HIRES spectra to search for RV trends from a potential third stellar-mass or substellar mass companion at long periods. In Section \ref{sec:method}, we present a comprehensive analysis of spectra, primary stellar RVs, and secondary stellar RVs for both systems. In Section \ref{sec:MassConstraints} we derive upper limits on putative companions at large separations. We conclude with a discussion of how future observations can improve our characterization of the architectures of Kepler-34 and Kepler-34 in Section \ref{sec:disc}.

\section{Methodology}
\label{sec:method}
\subsection{Radial Velocity Measurements}

Our analysis makes use of 22 previously obtained spectra of Kepler-34 and 13 spectra of Kepler-35 \citep{Welsh+12}. Each spectrum yields radial velocities for both the primary and secondary star. For the Kepler-34 spectra, four come from the HIRES spectrograph on Keck-1, eleven are from the Tull Coude spectrograph on the 2.7m Harlan J. Smith Telescope, and seven are from the HRS spectrograph on the Hobby-Eberly Telescope. For the Kepler-35 spectra, seven come from the HIRES spectrograph, three are from the FIES spectrograph on the Nordic Optical telescope, and three are from HRS. These observations all occurred between September and October in the year 2011.


\begin{figure*}
    \centering
    \subfigure[2011-09-02 09:38:57 UT\label{fig:sub1}]{%
        \includegraphics[width=0.3\textwidth]{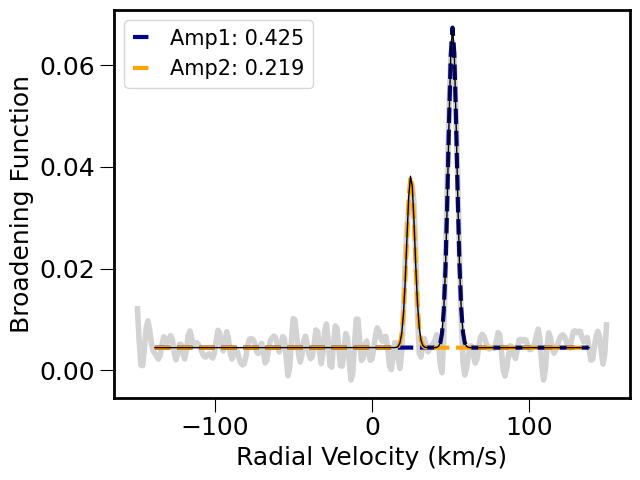}
    }
    \hfill
    \subfigure[2011-09-05 12:11:10 UT\label{fig:sub2}]{%
        \includegraphics[width=0.3\textwidth]{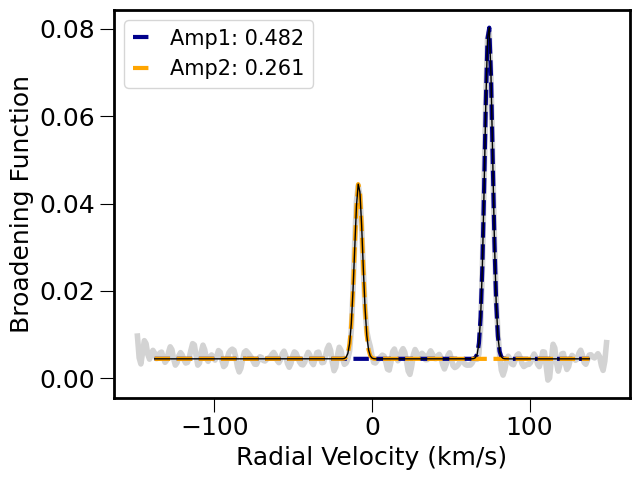}
    }
    \hfill
    \subfigure[2011-09-06 11:58:15 UT\label{fig:sub3}]{%
        \includegraphics[width=0.3\textwidth]{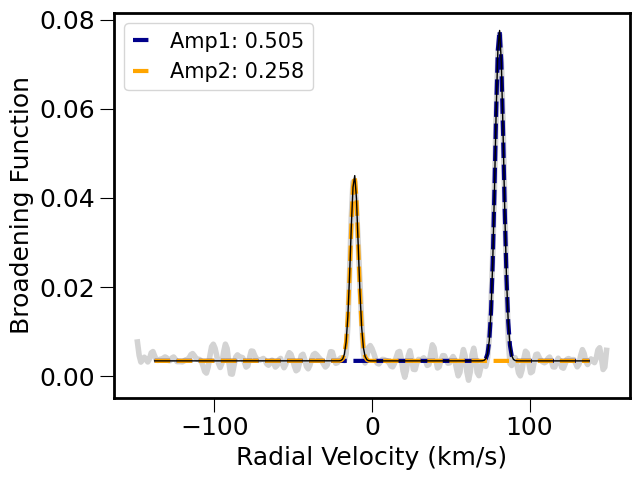}
    }

    \vspace{0.5cm}

    \subfigure[2011-09-10 08:00:05 UT\label{fig:sub4}]{%
        \includegraphics[width=0.3\textwidth]{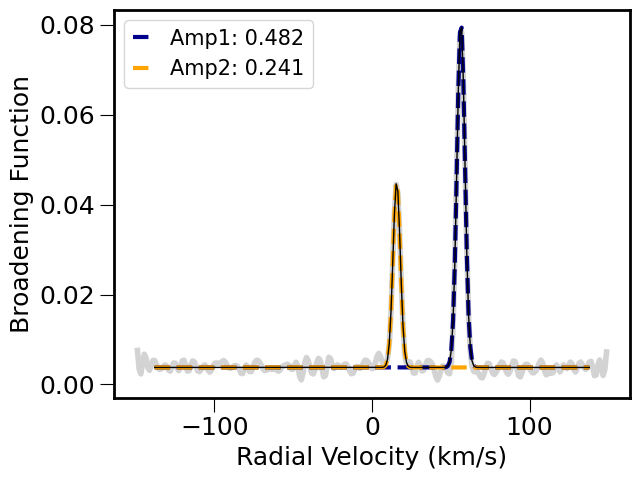}
    }
    \hfill
    \subfigure[2011-10-09 08:40:49 UT\label{fig:sub5}]{%
        \includegraphics[width=0.3\textwidth]{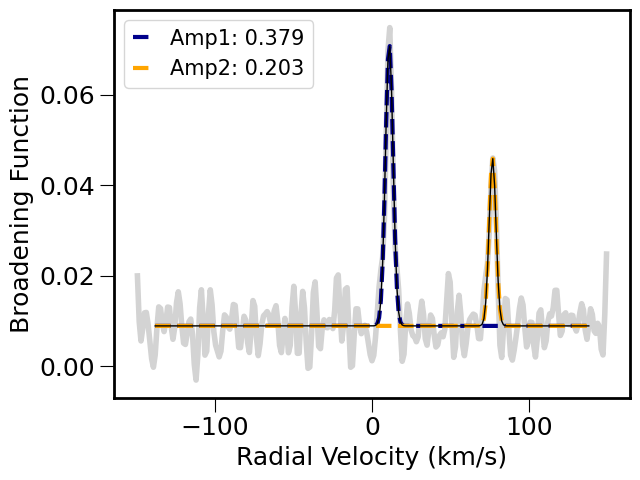}
    }
    \hfill
    \subfigure[2011-10-16 06:51:55 UT\label{fig:sub6}]{%
        \includegraphics[width=0.3\textwidth]{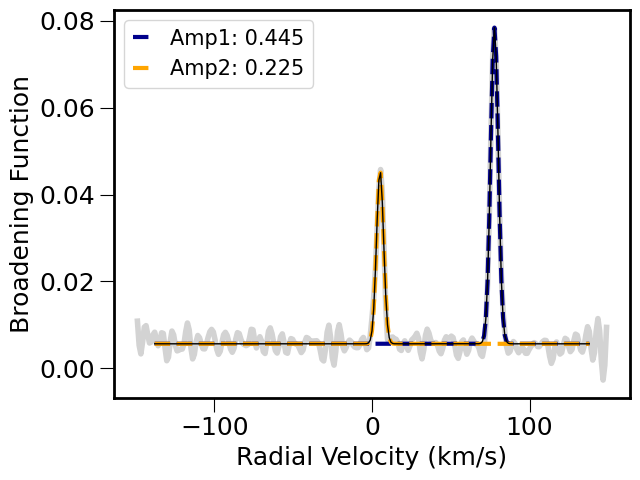}
    }

    \vspace{0.5cm}

    \subfigure[2011-10-17 07:56:14 UT\label{fig:sub7}]{%
        \includegraphics[width=0.3\textwidth]{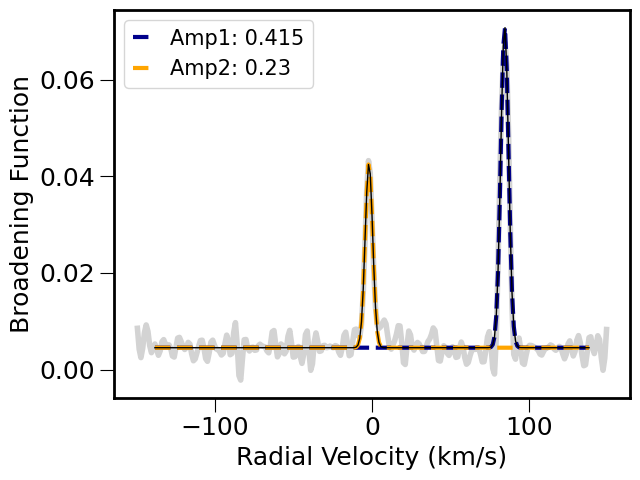}
    }
    \hfill
    \subfigure[2021-08-30 10:17:34 UT\label{fig:sub8}]{%
        \includegraphics[width=0.3\textwidth]{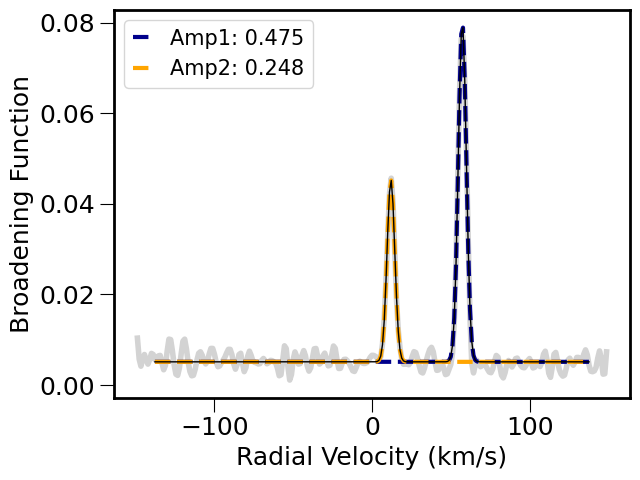}
    }
    \hfill
    \subfigure{\includegraphics[width=0.3\textwidth]{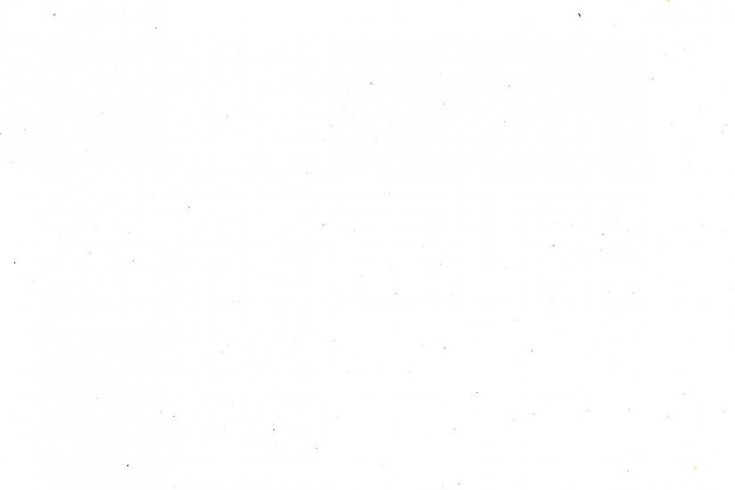}}

    \caption{Broadening Function of the HIRES spectra of Kepler-35 obtained in eight unique observations. Gaussian profiles are fitted to the peaks in the Broadening Function at the radial velocities associated with the primary star (blue) and secondary star (orange). The  area underneath the Gaussian profiles for each star are provided in the legend.}
    \label{fig:broadfunc_kep_35}
\end{figure*}

We obtained two new spectra of Kepler-34 in 2019 and 2021, and one of Kepler-35 in 2021, from the HIRES spectrograph on Keck-1. To measure the RVs of both stellar components on those dates, we used the broadening function (BF) technique in the publicly available code, {\tt SAPHIRES} \citep{Rucinski2002, Tofflemire+19}. To ensure consistency between the new and old HIRES RVs, we also reanalyzed the previously obtained HIRES spectra from 2011. We prepared the spectra for the BF analysis by trimming the edges from each order and resampled the spectra to a logarithmically spaced wavelength scale. We selected HD 182488 as our template star with a radial velocity of $-21.508$~km/s, for consistency with \cite{Welsh+12}. The BF is calculated separately for each order and subsequently combined into a single BF. This combined BF is then weighted according to the standard deviation of the BF edges, where there is no stellar signal. The uncertainty in the BF profile is assessed using a bootstrap Monte Carlo method, which generates 10,000 combined BFs through random sampling with replacement of the BF values determined from individual orders. 

Figures \ref{fig:broadfunc_kep_34} and \ref{fig:broadfunc_kep_35} show the {\tt SAPHIRES}-determined BF for all six and eight HIRES spectra of Kepler-34 and Kepler-35. We fit Gaussian profiles to the peaks of the BF to determine the observatory-frame radial velocities of the primary (colored blue) and secondary star (colored orange). We then applied the appropriate barycentric radial velocity corrections to obtain the RV time series of each star in the barycentric frame of reference.

The HIRES RVs obtained for Kepler-34 in 2019 have atypically large uncertainties, and a large RV residual. As seen in Figure \ref{fig:broadfunc_kep_34} in panel (e), only a singular peak was unambiguously identified from the the spectrum's broadening function. Our analysis of this specific observation reveals that the time of observation coincides with the secondary eclipse of Kepler-34. Given that the secondary eclipse has a long duration, the velocities are likely highly affected by the Rossiter-Mclaughlin effect. Although a secondary peak is weakly identified by {\tt SAPHIRES}, the centroid location of this peak varied signifiacntly across the various echelle orders that contributed to the BF. It's more likely that the second peak is noise and that both stellar components are nearly perfectly blended, making a precise RV measurement of both components challenging. We report the measured RV values on this date in Table \ref{tab:rvs}, but because of the poor quality of the spectrum and the blended nature of the BF peaks, we exclude this date in all further analysis. 

\begin{figure*}
    \centering
    \includegraphics[width=18cm]{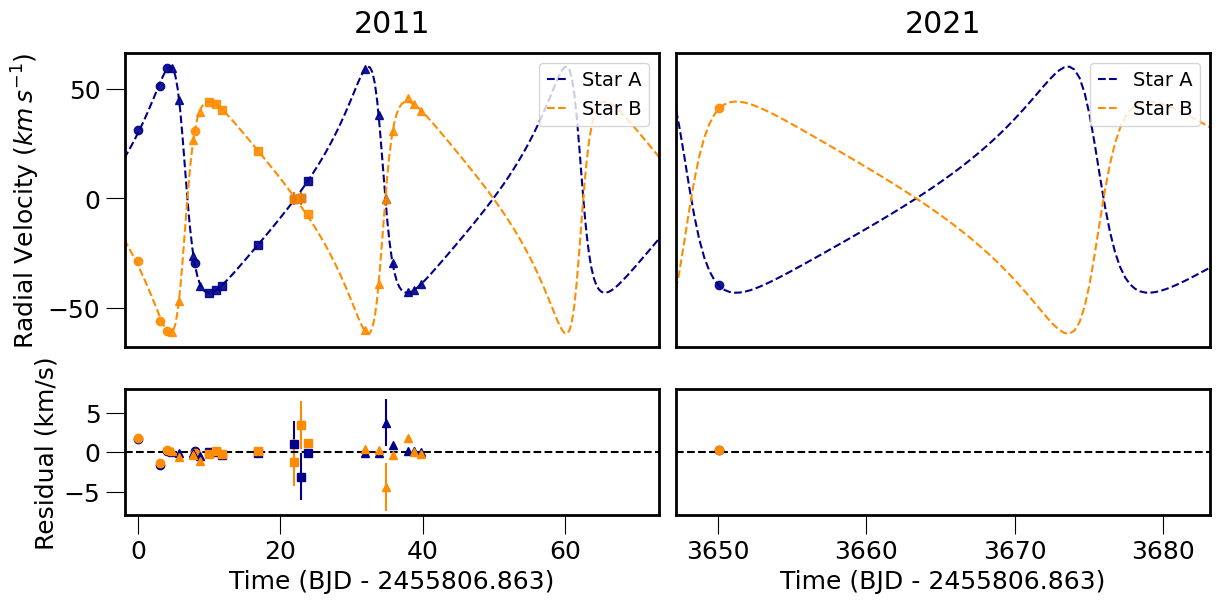}
    \caption{Top panels: radial velocities for Kepler-34 spanning over a decade (solid points). The $1 \sigma$ error bars are typically smaller than the point size. The Keplerian orbital parameters to generate the model of the RVs are listed in Table \ref{tab:bestfitvalues}. Note the significant temporal gap between the 2011 observations (left) and 2021 observation (right).  The RV measurements from the low-quality observation in 2019 are not shown, but were consistent with the lack of RV trend.  Bottom panels: The residuals. Circular markers denote Keck HIRES RVs, square markers denote HET HRS RVs, and triangle markers denote HJST Tull RVs.}
    \label{fig:kepler_34}
\end{figure*}

\subsection{Radial Velocity Curve Modeling}

\citet{Welsh+12} obtained precise measurements of the stellar and planetary orbital parameters for Kepler-34 and Kepler-35 by combining stellar spectra, transit timing variations, and photo-dynamical modeling of the systems. 

Initially we adopted all orbital parameters reported in \citet{Welsh+12} to model the RV curves of the primary and secondary star in each system with {\tt Radvel} \citep{Fulton2018}. The times of periastron were not reported in the main tables of the paper, but were included in the captions of Figures 1 and 2.  However, adopting the reported times of periastron resulted in large residual structures to the RVs.  We later determined that the times reported in Figures 1 and 2 of \citet{Welsh+12} are actually the times of conjunction (primary mid-eclipse times), not the times of periastron.  To correct for this error, we independently re-determined the best-fitting time of periastron for Kepler-34 and Kepler-35 from the RVs, finding $T_p = 2455007.125$ days and $T_p = 2455007.161$ days.


 
We introduce a constant offset to correct for zero-point velocity offsets between the spectrographs, based only on the RVs from 2011. We also introduce a free parameter, $\dot{\gamma}$, that captures long-term trends in the RVs potentially introduced by a stellar or substellar companion at orbital periods longer than the baseline of the RV measurements. 

\begin{figure*}[t]
    \centering
    \includegraphics[width=18cm]{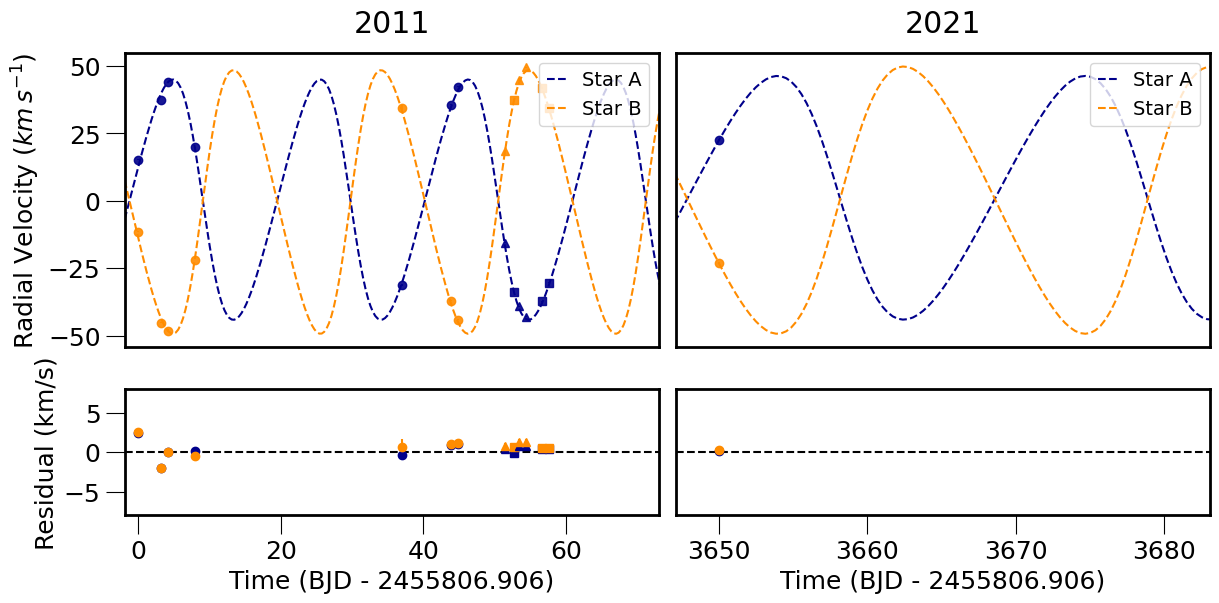}
    \caption{ Top panels: Same as Fig. \ref{fig:kepler_34}, but for Kepler-35. Bottom panels: The residuals. Circular markers denote Keck HIRES RVs, square markers denote HET HRS RVs, and triangle markers denote NOT FIES RVs.}
    \label{fig:kepler_35}
\end{figure*}

We fit for zero-point velocity offsets for each spectrograph, and a single value of $\dot{\gamma}$ fixed to zero, by maximizing the log-likelihood function:
\begin{equation}
    log(L) = -0.5 \left(\sum_i \frac{(RV_{i} - RV_{model}(t_i))^2 }{\sigma_{RV,i}^2}  + log(\sigma_{RV,i}^2) \right)
\end{equation}
where $RV_i$ is the $i$-th RV measurement, $RV_{model}(t_i)$ is the two-body keplerian modeled RV at time $t_i$ and $\sigma_{RV,i}$ is the uncertainity on the $i$-th RV measurement.
The modeled RV curves of Kepler-34 and Kepler-35 are generated using {\tt Radvel} with the orbital parameters from Table \ref{tab:bestfitvalues}. Figures \ref{fig:kepler_34} and \ref{fig:kepler_35} display the  RV curves and associated RV measurements for Kepler-34 and Kepler-35. The reanalyzed HIRES RVs are consistent with the values previously obtained by \citet{Welsh+12}.


\begin{table}
\centering
\begin{tabular}{ccc}
\toprule
{} & \multicolumn{2}{c}{Systems} \\
\cline{2-3}
{Orbital parameters} &      Kepler-34 &    Kepler-35\\

\midrule
\midrule

Mass of Star A, $M_A$ ($M_\odot$) &    1.0479 &  0.8877
\\

Mass of Star B, $M_B$, ($M_\odot$)  &    1.0208 &  0.8094
\\

Period, $P$ (days)  &    27.7958103 &  20.733666
\\

Eccentricity, $e$  &    0.52087 &  0.1421
\\

Inclination, $i$, (degrees)  &    89.8584 &  90.4238
\\

Semi-major axis, $a$ (AU) &    0.22882 &  0.17617
\\

Time of Periastron, (days) $T_p$  &    2455007.125 &  2455007.161
\\

Argument of Periapsis, $\omega$  (radians) &    1.2468  &  1.5058
\\

\bottomrule
\end{tabular}
\caption{The orbital parameters input into {\tt Radvel} for the two-body Keplerian models of Kepler-34 and Kepler-35.}
\label{tab:bestfitvalues}
\end{table}

\subsection{Lomb-Scargle Periodogram of the RV Residuals}
To search for periodic signals in the RVs at orbital periods commensurate with our observational baseline and shorter, we computed a Lomb-Scargle periodogram of the RV residuals. This analysis reveals a lack of a significant peak detection that can be indicative of a companion. This outcome was anticipated, given the incorporation of only one additional RV measurement over our 10-year baseline. A more meaningful Lomb-Scargle periodogram could be obtained with an increased number of RV observations in the future. 

\begin{figure*}[t]
    \centering
    \includegraphics[width=18cm]{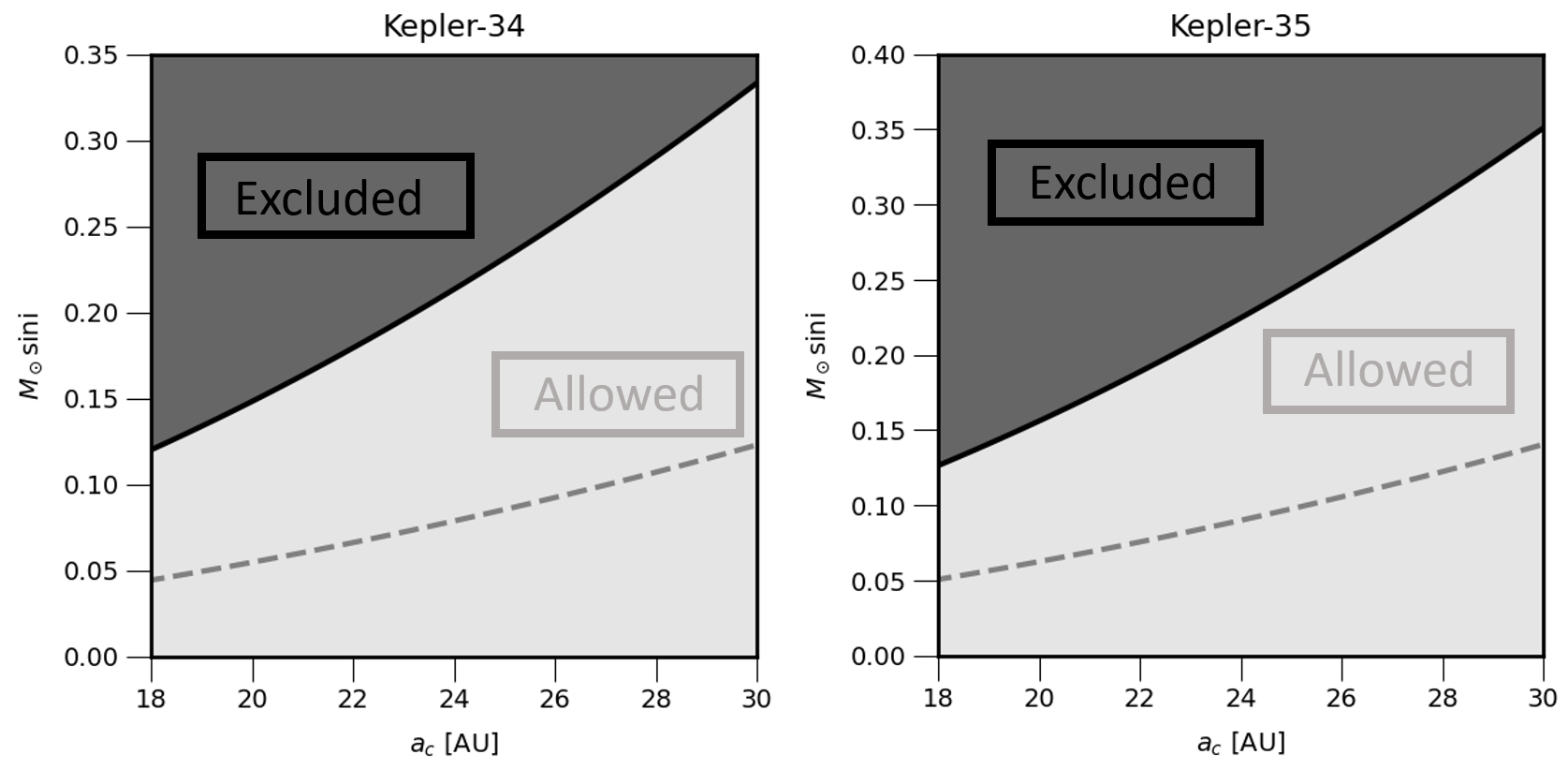}
    \caption{Upper limits on a proposed companion in the Kepler-34 and Kepler-35 system as a function of companion mass and semi-major axis. The dashed grey curve is obtained from the mean value of $\dot{\gamma}$ and the light grey region is the $\pm 3\sigma$ interval. The dark grey region corresponds to masses above our $3\sigma$ upper limit.}
    \label{fig:mass_constraint}
\end{figure*}

\subsection{Modeling Long-Term Trends in the RVs}
Companions with orbital periods significantly longer than the baseline scale were searched for by modeling a constant acceleration acting on the barycenter of the binary. We fixed the zero-point velocity offsets for all RV measurements in our data, allowing $\dot{\gamma}$ to be the only remaining free parameter. This ensures that any deviation of the most recently obtained RVs from the modeled RV curves can be attributed to a dynamical process from the system. A significant detection of $\dot{\gamma}$ could potentially correspond to a companion in the system with degeneracy between the semi-major axis and mass. A non-detection of $\dot{\gamma}$ can be used to place upper limits of putative, hidden companions. We estimated the uncertainty in our parameters with a Markov-Chain Monte Carlo analysis (MCMC), implemented via the python package {\tt emcee} \citep{Mackey+13}.  Our uncertainty in  $\dot{\gamma}$ is driven by the uncertainty on the individual RV measurements, which have a root-mean square error (RMSE) of 1\,km/s with respect to the best fit (Table \ref{tab:kepsystems}).

\section{Mass-Period Constraints on Companions}
\label{sec:MassConstraints}

\begin{table}[b]
\centering
\begin{tabular}{ccc}
\toprule
{} & \multicolumn{2}{c}{Systems} \\
\cline{2-3}
{RV Trends and Errors} &      Kepler-34 &    Kepler-35\\

\midrule
\midrule
$\dot{\gamma}$            &    $0.00007^{+0.00004}_{-0.00004}$   & $0.00008^{+0.00004}_{-0.00004}$   \\
RMSE of Star A            &    1.17 km/s &  0.99 km/s   \\
RMSE of Star B            &    1.40 km/s &  1.16 km/s   \\
RMSE of Star A [HIRES]    &    1.04 km/s &  1.23 km/s \\
RMSE of Star B [HIRES]    &    1.05 km/s &  1.32 km/s \\
\bottomrule
\end{tabular}
\caption{Radial velocity trends and errors for Kepler-34 and Kepler-35. }
\label{tab:kepsystems}
\end{table}

As a proof of concept, we assume a companion on a co-planar, circular orbit around the barycenter of the binary stars. Then, the mass of that companion can be calculated by dividing the gravitational force by its acceleration on the barycenter of the binary system, yielding:
\begin{equation}
    M_{c} \sin(i) = \frac{\dot{\gamma}a_{c}^{2}}{G}
\end{equation}
where $a_c$ is the semi-major axis of the companion object, $i$ is the inclination, and G is the gravitational constant. We are restricted to companions on orbital separations much greater than the span of the RV measurements (i.e., $a_{c}>15 AU)$. Possible companions with orbital periods on a scale of less than a decade will average the direction of the RV change, so this approach is only suitable for placing upper limits on masses at long periods.

\subsection{Kepler-34}
In Figure \ref{fig:mass_constraint}, we illustrate the conservative upper limit for the mass of a companion in the Kepler-34 system, derived from $\dot{\gamma}$. The mass limit obtained from the mean value of $\dot{\gamma}$ is shown as a dashed grey curve, while the shaded grey region represents the $\pm 3\sigma$ interval. Stellar companions with a $M \sin(i)$ exceeding 0.33 solar masses and situated within a semi-major axis of 30 AU are ruled out by the new RV data. However, M-type stars or substellar objects at 30 AU are consistent with the RV measurements.  


\subsection{Kepler-35}
In Figure \ref{fig:mass_constraint}, we also illustrate the conservative upper limit for the mass of a companion in the Kepler-35 system, derived from $\dot{\gamma}$.  Stellar companions with a $M \sin(i)$ exceeding 0.35 solar masses and situated within a semi-major axis of 30 AU are ruled out. 
Similar to Kepler-34, M-type stars or substellar objects are considered feasible candidates for companions to the binary stars from only the RV measurements. As is the case for Kepler-34, continual monitoring of the Kepler-35 system would provide a significant amount of RVs over a longer baseline to constrain the companion parameter space.

\section{Discussion and Conclusion}
\label{sec:disc}

Here we search for RV trends over a decade long baseline in the Kepler-34 and Kepler-35 systems that may be indicative of tertiary companions. We reanalyze both the previous and newly obtained HIRES RVs which are consistent with the orbital fits from \citep{Welsh+12}. Given this agreement, we constrain a mass upper limit for potential third stellar-mass (or substellar mass) companions on circular orbits. For Kepler-34, a tertiary $M \sin(i)$ upper limit ranges from 0.12 $M_\odot$ located at a radius of 18AU to 0.33 $M_\odot$ at 30AU from the barycenter of the stellar binary (3$\sigma$ conf.). For Kepler-35, a tertiary $M \sin(i)$ upper limit ranges from 0.13 $M_\odot$ located at a radius of 18AU to 0.35 $M_\odot$ at 30AU from the barycenter of the stellar binary (3$\sigma$ conf.).

The major factor limiting our ability to decrease the mass upper limit to substellar and planetary masses is errors in the RVs of order 0.1\,km/s that arise from the challenges of characterizing double-lined spectra. This challenge has been noted in previous studies of double-lined binaries, \citep{Konacki2009, Konacki2010}. In contrast, spectroscopic studies that have focused on single-lined, low-mass eclipsing binaries \citep[BEBOP;][]{Martin2019+}, and are sensitive to Saturn-mass circumbinary planets within orbital periods of 1000 days \citep{Standing2022}, leading to the RV detection of Kepler-16 b \citep{Triaud2022} and TOI-1338/BEBOP-1 c\citep{Standing2023}.

Several advancements are promising for improving the mass sensitivity in the search for third objects around double-lined spectroscopic binaries.  Extreme-precision RV (EPRV) instruments include hardware advancements that yield an ultra-stable point-spread function (PSF) of the starlight entering the spectrograph. Several examples include the Keck Planet Finder \citep{Gibson2018}, iLocator \citep{Crepp2016+}, EXPRES \citep{Jurgenson2016+, Petersburg2020+}, and NEID \citep{Schwab2016+}.  These instruments all are temperature-stabilized to within a few mK, vacuum-sealed, and fed by optical fibers that scramble and/or reduce the modal noise of the PSF.  These designs will reduce the errors that arise from changes in the stellar PSF.  Meanwhile, \citet{Sairam2024} used Gaussian Processes to detect a circumbinary planet in the TIC 172900988 system with spectra obtained from the SOPHIE spectrograph, demonstrating that it might be possible to improve the RV characterization of both new and archival spectra.  These advancements suggest a bright future for the characterization of circumbinary planets and their host star environments.


\section*{Acknowledgements}
This material was carried out at The University of Notre Dame, with support from the National Science Foundation REU Program (grant PHY-2050527). L.M.W.\ acknowledges support from the NASA Exoplanet Research Program through grant 80NSSC23K0269 and from NASA-Keck Key Stragetic Mission Support Grant No. 80NSSC19K1475.  
\bibliography{references}

\begin{thebibliography}{}
\expandafter\ifx\csname natexlab\endcsname\relax\def\natexlab#1{#1}\fi
\providecommand{\url}[1]{\href{#1}{#1}}
\providecommand{\dodoi}[1]{doi:~\href{http://doi.org/#1}{\nolinkurl{#1}}}
\providecommand{\doeprint}[1]{\href{http://ascl.net/#1}{\nolinkurl{http://ascl.net/#1}}}
\providecommand{\doarXiv}[1]{\href{https://arxiv.org/abs/#1}{\nolinkurl{https://arxiv.org/abs/#1}}}

\bibitem[{{Armstrong} {et~al.}(2014){Armstrong}, {Osborn}, {Brown}, {Faedi}, {G{\'o}mez Maqueo Chew}, {Martin}, {Pollacco}, \& {Udry}}]{Armstrong2014}
{Armstrong}, D.~J., {Osborn}, H.~P., {Brown}, D.~J.~A., {et~al.} 2014, \mnras, 444, 1873, \dodoi{10.1093/mnras/stu1570}

\bibitem[{{Crepp} {et~al.}(2016){Crepp}, {Crass}, {King}, {Bechter}, {Bechter}, {Ketterer}, {Reynolds}, {Hinz}, {Kopon}, {Cavalieri}, {Fantano}, {Koca}, {Onuma}, {Stapelfeldt}, {Thomes}, {Wall}, {Macenka}, {McGuire}, {Korniski}, {Zugby}, {Eisner}, {Gaudi}, {Hearty}, {Kratter}, {Kuchner}, {Micela}, {Nelson}, {Pagano}, {Quirrenbach}, {Schwab}, {Skrutskie}, {Sozzetti}, {Woodward}, \& {Zhao}}]{Crepp2016+}
{Crepp}, J.~R., {Crass}, J., {King}, D., {et~al.} 2016, in Society of Photo-Optical Instrumentation Engineers (SPIE) Conference Series, Vol. 9908, Ground-based and Airborne Instrumentation for Astronomy VI, ed. C.~J. {Evans}, L.~{Simard}, \& H.~{Takami}, 990819, \dodoi{10.1117/12.2233135}

\bibitem[{{Doyle} {et~al.}(2011){Doyle}, {Carter}, {Fabrycky}, {Slawson}, {Howell}, {Winn}, {Orosz}, {P{\v{r}}sa}, {Welsh}, {Quinn}, {Latham}, {Torres}, {Buchhave}, {Marcy}, {Fortney}, {Shporer}, {Ford}, {Lissauer}, {Ragozzine}, {Rucker}, {Batalha}, {Jenkins}, {Borucki}, {Koch}, {Middour}, {Hall}, {McCauliff}, {Fanelli}, {Quintana}, {Holman}, {Caldwell}, {Still}, {Stefanik}, {Brown}, {Esquerdo}, {Tang}, {Furesz}, {Geary}, {Berlind}, {Calkins}, {Short}, {Steffen}, {Sasselov}, {Dunham}, {Cochran}, {Boss}, {Haas}, {Buzasi}, \& {Fischer}}]{Doyle2011}
{Doyle}, L.~R., {Carter}, J.~A., {Fabrycky}, D.~C., {et~al.} 2011, Science, 333, 1602, \dodoi{10.1126/science.1210923}

\bibitem[{{Foreman-Mackey} {et~al.}(2013){Foreman-Mackey}, {Hogg}, {Lang}, \& {Goodman}}]{Mackey+13}
{Foreman-Mackey}, D., {Hogg}, D.~W., {Lang}, D., \& {Goodman}, J. 2013, \pasp, 125, 306, \dodoi{10.1086/670067}

\bibitem[{{Fulton} {et~al.}(2018){Fulton}, {Petigura}, {Blunt}, \& {Sinukoff}}]{Fulton2018}
{Fulton}, B.~J., {Petigura}, E.~A., {Blunt}, S., \& {Sinukoff}, E. 2018, \pasp, 130, 044504, \dodoi{10.1088/1538-3873/aaaaa8}

\bibitem[{{Gibson} {et~al.}(2018){Gibson}, {Howard}, {Roy}, {Smith}, {Halverson}, {Edelstein}, {Kassis}, {Wishnow}, {Raffanti}, {Allen}, {Chin}, {Coutts}, {Cowley}, {Curtis}, {Deich}, {Feger}, {Finstad}, {Gurevich}, {Ishikawa}, {James}, {Jhoti}, {Lanclos}, {Lilley}, {Miller}, {Milner}, {Payne}, {Rider}, {Rockosi}, {Sandford}, {Schwab}, {Seifahrt}, {Sirk}, {Smith}, {Stuermer}, {Weisfeiler}, {Wilcox}, {Vandenberg}, \& {Wizinowich}}]{Gibson2018}
{Gibson}, S.~R., {Howard}, A.~W., {Roy}, A., {et~al.} 2018, in Society of Photo-Optical Instrumentation Engineers (SPIE) Conference Series, Vol. 10702, Ground-based and Airborne Instrumentation for Astronomy VII, ed. C.~J. {Evans}, L.~{Simard}, \& H.~{Takami}, 107025X, \dodoi{10.1117/12.2311565}

\bibitem[{Jurgenson {et~al.}(2016)Jurgenson, Fischer, McCracken, Sawyer, Szymkowiak, Davis, Muller, \& Santoro}]{Jurgenson2016+}
Jurgenson, C., Fischer, D., McCracken, T., {et~al.} 2016, in Ground-based and Airborne Instrumentation for Astronomy VI, ed. C.~J. Evans, L.~Simard, \& H.~Takami, Vol. 9908, International Society for Optics and Photonics (SPIE), 99086T, \dodoi{10.1117/12.2233002}

\bibitem[{{Konacki} {et~al.}(2009){Konacki}, {Muterspaugh}, {Kulkarni}, \& {He{\l}miniak}}]{Konacki2009}
{Konacki}, M., {Muterspaugh}, M.~W., {Kulkarni}, S.~R., \& {He{\l}miniak}, K.~G. 2009, \apj, 704, 513, \dodoi{10.1088/0004-637X/704/1/513}

\bibitem[{{Konacki} {et~al.}(2010){Konacki}, {Muterspaugh}, {Kulkarni}, \& {He{\l}miniak}}]{Konacki2010}
---. 2010, \apj, 719, 1293, \dodoi{10.1088/0004-637X/719/2/1293}

\bibitem[{{Li} {et~al.}(2016){Li}, {Holman}, \& {Tao}}]{Li2016}
{Li}, G., {Holman}, M.~J., \& {Tao}, M. 2016, \apj, 831, 96, \dodoi{10.3847/0004-637X/831/1/96}

\bibitem[{{Marcy} \& {Butler}(1992)}]{Marcy1992}
{Marcy}, G.~W., \& {Butler}, R.~P. 1992, \pasp, 104, 270, \dodoi{10.1086/132989}

\bibitem[{{Martin} \& {Triaud}(2014)}]{Martin2014}
{Martin}, D.~V., \& {Triaud}, A. H.~M.~J. 2014, \aap, 570, A91, \dodoi{10.1051/0004-6361/201323112}

\bibitem[{{Martin} {et~al.}(2019){Martin}, {Triaud}, {Udry}, {Marmier}, {Maxted}, {Collier Cameron}, {Hellier}, {Pepe}, {Pollacco}, {S{\'e}gransan}, \& {West}}]{Martin2019+}
{Martin}, D.~V., {Triaud}, A. H.~M.~J., {Udry}, S., {et~al.} 2019, \aap, 624, A68, \dodoi{10.1051/0004-6361/201833669}

\bibitem[{{Petersburg} {et~al.}(2020){Petersburg}, {Ong}, {Zhao}, {Blackman}, {Brewer}, {Buchhave}, {Cabot}, {Davis}, {Jurgenson}, {Leet}, {McCracken}, {Sawyer}, {Sharov}, {Tronsgaard}, {Szymkowiak}, \& {Fischer}}]{Petersburg2020+}
{Petersburg}, R.~R., {Ong}, J.~M.~J., {Zhao}, L.~L., {et~al.} 2020, \aj, 159, 187, \dodoi{10.3847/1538-3881/ab7e31}

\bibitem[{Rucinski(2002)}]{Rucinski2002}
Rucinski, S.~M. 2002, The Astronomical Journal, 124, 1746, \dodoi{10.1086/342342}

\bibitem[{{Sairam} {et~al.}(2024){Sairam}, {Triaud}, {Baycroft}, {Orosz}, {Boisse}, {Heidari}, {Sebastian}, {Dransfield}, {Martin}, {Santerne}, \& {Standing}}]{Sairam2024}
{Sairam}, L., {Triaud}, A. H.~M.~J., {Baycroft}, T.~A., {et~al.} 2024, \mnras, 527, 2261, \dodoi{10.1093/mnras/stad3136}

\bibitem[{{Schwab} {et~al.}(2016){Schwab}, {Rakich}, {Gong}, {Mahadevan}, {Halverson}, {Roy}, {Terrien}, {Robertson}, {Hearty}, {Levi}, {Monson}, {Wright}, {McElwain}, {Bender}, {Blake}, {St{\"u}rmer}, {Gurevich}, {Chakraborty}, \& {Ramsey}}]{Schwab2016+}
{Schwab}, C., {Rakich}, A., {Gong}, Q., {et~al.} 2016, in Society of Photo-Optical Instrumentation Engineers (SPIE) Conference Series, Vol. 9908, Ground-based and Airborne Instrumentation for Astronomy VI, ed. C.~J. {Evans}, L.~{Simard}, \& H.~{Takami}, 99087H, \dodoi{10.1117/12.2234411}

\bibitem[{{Standing} {et~al.}(2022){Standing}, {Triaud}, {Faria}, {Martin}, {Boisse}, {Correia}, {Deleuil}, {Dransfield}, {Gillon}, {H{\'e}brard}, {Hellier}, {Kunovac}, {Maxted}, {Mardling}, {Santerne}, {Sairam}, \& {Udry}}]{Standing2022}
{Standing}, M.~R., {Triaud}, A. H.~M.~J., {Faria}, J.~P., {et~al.} 2022, \mnras, 511, 3571, \dodoi{10.1093/mnras/stac113}

\bibitem[{{Standing} {et~al.}(2023){Standing}, {Sairam}, {Martin}, {Triaud}, {Correia}, {Coleman}, {Baycroft}, {Kunovac}, {Boisse}, {Cameron}, {Dransfield}, {Faria}, {Gillon}, {Hara}, {Hellier}, {Howard}, {Lane}, {Mardling}, {Maxted}, {Miller}, {Nelson}, {Orosz}, {Pepe}, {Santerne}, {Sebastian}, {Udry}, \& {Welsh}}]{Standing2023}
{Standing}, M.~R., {Sairam}, L., {Martin}, D.~V., {et~al.} 2023, Nature Astronomy, 7, 702, \dodoi{10.1038/s41550-023-01948-4}

\bibitem[{Tofflemire(2019)}]{Tofflemire+19}
Tofflemire, B.~M. 2019, tofflemire/saphires: Zenodo archive, 0.1.11,  Zenodo, \dodoi{10.5281/zenodo.3497509}

\bibitem[{{Triaud} {et~al.}(2022){Triaud}, {Standing}, {Heidari}, {Martin}, {Boisse}, {Santerne}, {Correia}, {Acu{\~n}a}, {Battley}, {Bonfils}, {Carmona}, {Collier Cameron}, {Cort{\'e}s-Zuleta}, {Dransfield}, {Dalal}, {Deleuil}, {Delfosse}, {Faria}, {Forveille}, {Hara}, {H{\'e}brard}, {Hoyer}, {Kiefer}, {Kunovac}, {Maxted}, {Martioli}, {Miller}, {Nelson}, {Poveda}, {Rein}, {Sairam}, {Udry}, \& {Willett}}]{Triaud2022}
{Triaud}, A. H.~M.~J., {Standing}, M.~R., {Heidari}, N., {et~al.} 2022, \mnras, 511, 3561, \dodoi{10.1093/mnras/stab3712}

\bibitem[{{Welsh} {et~al.}(2012){Welsh}, {Orosz}, {Carter}, {Fabrycky}, {Ford}, {Lissauer}, {Pr{\v{s}}a}, {Quinn}, {Ragozzine}, {Short}, {Torres}, {Winn}, {Doyle}, {Barclay}, {Batalha}, {Bloemen}, {Brugamyer}, {Buchhave}, {Caldwell}, {Caldwell}, {Christiansen}, {Ciardi}, {Cochran}, {Endl}, {Fortney}, {Gautier}, {Gilliland}, {Haas}, {Hall}, {Holman}, {Howard}, {Howell}, {Isaacson}, {Jenkins}, {Klaus}, {Latham}, {Li}, {Marcy}, {Mazeh}, {Quintana}, {Robertson}, {Shporer}, {Steffen}, {Windmiller}, {Koch}, \& {Borucki}}]{Welsh+12}
{Welsh}, W.~F., {Orosz}, J.~A., {Carter}, J.~A., {et~al.} 2012, \nat, 481, 475, \dodoi{10.1038/nature10768}

\end{thebibliography}

\section{Appendix}

\begin{table}[h]
\centering
\resizebox{\linewidth}{!}{%
\begin{tabular}{|l|l|l|l|l|l|} 
\hline
Date       & BJD & $RV_A$     & $RV_B$     & Telescope/  & Flag    \\
YYYY-MM-DD & (2,400,000+)  & (km\,s$^{-1
}$) & (km\,s$^{-1}$) & Spectrograph           &         \\
\hline
2011-09-02 &  55806.862608    & 35.129 $\pm$ 0.0683          & -25.236 $\pm$ 0.0934         & Keck HIRES &         \\
\hline
2011-09-05 & 55810.004147    & 55.253 $\pm$ 0.0553 &  -52.495 $\pm$ 0.0943 & Keck HIRES &         \\
\hline
2011-09-06 & 55810.996812    & 63.306 $\pm$ 0.0670 & -57.188 $\pm$ 0.0844         & Keck HIRES &         \\
\hline
2011-09-07 & 55811.6711428    & 65.097 $\pm$  0.165           & -56.06 $\pm$ 0.174          & HJST TULL  &         \\
\hline
2011-09-08 & 55812.6440094    & 50.196 $\pm$  0.183           & -41.578 $\pm$ 0.222          & HJST TULL  &         \\
\hline
2011-09-10 & 55814.6489276    & -21.195 $\pm$ 0.164           &  31.873 $\pm$ 0.207          & HJST TULL  &         \\
\hline
2011-09-10 & 55814.831011    &  -26.138 $\pm$ 0.0515 & 34.299 $\pm$ 0.0595 & KECK HIRES &         \\
\hline
2011-09-11 & 55815.6410401    & -34.959 $\pm$ 0.189           & 44.982 $\pm$ 0.254           & HJST TULL  &         \\
\hline
2011-09-12 & 55816.7766219    & -39.052 $\pm$ 0.3             & 48.293 $\pm$ 0.275          & HET HRS    &         \\
\hline
2011-09-13 & 55817.7719168     & -37.806 $\pm$ 0.155           & 47.505 $\pm$ 0.184          & HET HRS    &         \\
\hline
2011-09-14 & 55818.7297401      & -35.704 $\pm$ 0.2             & 44.505 $\pm$ 0.255          & HET HRS    &         \\
\hline
2011-09-19 & 55823.7296067      & -17.16 $\pm$ 0.075            & 26.103 $\pm$ 0.086          & HET HRS    &         \\
\hline
2011-09-24 & 55828.7105309     & 4.077 $\pm$ 3.000                 & 4.077 $\pm$ 3.000          & HET HRS    &         \\
\hline
2011-09-25 & 55829.6967155    & 4.309 $\pm$ 3.000                 & 4.309 $\pm$ 3.000          & HET HRS    &         \\
\hline
2011-09-26 & 55830.7056930   & 12.076 $\pm$ 0.090             & -2.822 $\pm$ 0.205          & HET HRS    &         \\
\hline
2011-10-04 & 55838.7132290   & 64.332 $\pm$ 0.129            & -55.151 $\pm$ 0.149          & HJST TULL  &         \\
\hline
2011-10-06 &  55840.6464517   & 43.603 $\pm$ 0.186            & -33.956 $\pm$ 0.198          & HJST TULL  &         \\
\hline
2011-10-7  & 55841.6997478    & 4.945 $\pm$ 3.000                 & 4.945 $\pm$ 3.000          & HJST TULL  &         \\
\hline
2011-10-8  & 55842.6488201    & -24.218 $\pm$ 0.253           & 36.07 $\pm$ 0.456          & HJST TULL  &         \\
\hline
2011-10-10 & 55844.7159536    & -37.802 $\pm$ 0.298           & 51.318 $\pm$ 0.287          & HJST TULL  &         \\
\hline
2011-10-11 & 55845.6407652    & -36.539 $\pm$ 0.205           & 48.268 $\pm$ 0.22          & HJST TULL  &         \\
\hline
2011-10-12 & 55846.6382723    & -33.893 $\pm$ 0.185           & 45.189 $\pm$ 0.232          & HJST TULL  &         \\
\hline
2019-12-15 & 58832.723406    & 11.706	$\pm$ 0.0872   & 11.706 $\pm$ 0.0872            & KECK HIRES & ! \\
\hline
2021-08-30 & 59456.929329    & -36.132 $\pm$ 0.211 &  44.948 $\pm$ 0.184 & KECK HIRES &         \\
\hline
\end{tabular}
}
\caption{Radial velocities for Kepler-34}
\label{tab:rvs}
\end{table}

\begin{table}
\centering
\resizebox{\linewidth}{!}{%
\begin{tabular}{|l|l|l|l|l|l|} 
\hline
Date       & BJD & $RV_A$     & $RV_B$     & Telescope/  & Flag    \\
YYYY-MM-DD & (2,400,000+)  & (km\,s$^{-1
}$) & (km\,s$^{-1}$) & Spectrograph           &         \\
\hline
2011-09-02 & 55806.906958                                              & 36.859 $\pm$ 0.166   & 10.176 $\pm$ 0.226   & Keck HIRES &      \\ \hline
2011-09-05 & 55810.012586                                              & 58.966 $\pm$ 0.085   & -23.5288 $\pm$ 0.179 & Keck HIRES &      \\ \hline
2011-09-06 & 55811.004419                                              & 65.542 $\pm$ 0.086   & -26.487 $\pm$ 0.439  & Keck HIRES &      \\ \hline
2011-09-10 & 55814.838820                                             & 41.413 $\pm$ 0.059   & -0.332 $\pm$ 0.139   & Keck HIRES &      \\ \hline
2011-10-09 & 55843.864476                                              & -9.559 $\pm$ 0.493   & 56.098 $\pm$ 0.960   & Keck HIRES &      \\ \hline
2011-10-16 & 55850.788613                                              & 57.209 $\pm$ 0.189   & -15.385 $\pm$ 0.222  & Keck HIRES &      \\ \hline
2011-10-17 & 55851.833203                                              & 63.960 $\pm$ 0.064   & -22.450 $\pm$ 0.286  & Keck HIRES &      \\ \hline
2011-10-23 & 55858.3392335                                              & 7.137 $\pm$ 0.176    & 41.105 $\pm$ 0.352   & NOT FIES   &      \\ \hline
2011-10-25 & 55859.6190952                                              & -10.447 $\pm$ 0.093  & 60.846 $\pm$ 0.202   & HET HRS    &      \\ \hline
2011-10-25 & 55860.3451372                                              & -16.345 $\pm$ 0.224  & 67.745 $\pm$ 0.387   & NOT FIES   &      \\ \hline
2011-10-26 & 55861.3429667                                              & -20.278 $\pm$ 0.189  & 72.235 $\pm$ 0.334   & NOT FIES   &      \\ \hline
2011-10-29 & 55863.6344314                                              & -13.903 $\pm$ 0.131  & 65.182 $\pm$ 0.246   & HET HRS    &      \\ \hline
2011-10-30 & 55864.6017025                                              & -7.053 $\pm$ 0.128   & 57.578 $\pm$ 0.218   & HET HRS    &      \\ \hline
2021-08-30 & 59456.93268  & 43.35076 $\pm$ 0.210 & -1.82994 $\pm$ 0.219 & Keck HIRES &      \\ \hline
\end{tabular}
}
\caption{Radial Velocities for Kepler-35}
\label{tab:zeropoint}
\end{table}

\end{document}